# Measurement of n-resolved State-Selective Charge Exchange in Ne$^{(8;9)+}$ Collision with He and H$_2$

J. W. Xu,[1,2] C. X. Xu,[3,4] R. T. Zhang,[1,*] X. L. Zhu,[1,2,†] W. T. Feng,[1] L. Gu,[5,6] G. Y. Liang,[7] D. L. Guo,[1] Y. Gao,[1]
D. M. Zhao,[1] S. F. Zhang,[1,2] M. G. Su,[3,4] and X. Ma[1,2,‡]

[1] Institute of Modern Physics, Chinese Academy of Sciences, Lanzhou 730000, China
[2] University of Chinese Academy of Sciences, Beijing 100049, China
[3] Key Laboratory of Atomic and Molecular Physics & Functional Material of Gansu Province, College of Physics
and Electronic Engineering, Northwest Normal University, Lanzhou 730070, China
[4] Joint Laboratory of Atomic and Molecular Physics, Northwest Normal University and Institute of Modern Physics,
Chinese Academy of Sciences, Lanzhou 730070, China
[5] RIKEN High Energy Astrophysics Laboratory, 2-1 Hirosawa, Wako, Saitama 351-0198, Japan
[6] SRON Netherlands Institute for Space Research, Sorbonnelaan 2, 3584 CA Utrecht, the Netherlands
[7] Key Laboratory of Optical Astronomy, National Astronomical Observatories, Chinese Academy of Sciences,
Beijing 100012, China2

## Abstract

Charge exchange between highly charged ions and neutral atoms and molecules has been considered as one of important mechanisms controlling soft X-ray emissions in many astrophysical objects and environments. Whereas for modelling charge exchange soft X-ray emission these data of n and l-resolved state selective capture cross sections are often obtained by empirical and semi-classical theory calculations. With a newly-built cold target recoil ion momentum spectroscopy (COLTRIMS) apparatus, we perform a series of measurements of Ne$^{(8;9)+}$ ions charge exchange with He and H$_2$ for the collision energy ranging from 1 to 24.75 keV/u. n-resolved state-selective capture cross sections are reported. By comparing the measured state-selective capture cross sections to these calculated by the multichannel Landau-Zener method (MCLZ), it is found that MCLZ calculations are in a good agreement with the measurement for the dominant n capture for He target. Furthermore, by using $nl$-resolved cross sections calculated by MCLZ and applying l distributions commonly-used in astrophysical literatures to experimentally derived n-resolved cross sections, we calculate the soft X-ray emissions in the charge exchange between 4 keV/u Ne$^{8+}$ and He by considering the radiative cascade from the excited Ne$^{7+}$ ion. Reasonable agreement is found in comparison to the measurement for even and separable models, and MCLZ calculations give results in a better agreement.

---

* zhangrt@impcas.ac.cn
† zhuxiaolong@impcas.ac.cn
‡ x.ma@impcas.ac.cn

## 1. INTRODUCTIONS

Single electron capture (SEC) charge exchange (CX) between highly charged ions and neutral atoms or molecules is a process where one target electron is transferred onto an ion. It has long been conjectured as a possible source of astrophysical X-ray emission (Hayakawa 1960; Joseph & Gary 1969; Pravdo & Boldt 1975). Around 20 yr ago, the CX of solar wind ions with neutrals was confirmed to be responsible for the X-ray emission from the comas of comets and the diffuse soft X-ray background (Lisse et al. 1996; Cravens 1997; Cox 1998; Cravens 2000; Beiersdorfer et al. 2003; Dennerl et al. 2006; Branduardi-Raymont et al. 2007; Koutroumpa et al. 2009). Recently, important evidence for CX has been found in the spectra of supernova remnants (Katsuda et al. 2011), around star-forming galaxies (Liu et al. 2012), and in clusters of galaxies (Fabian et al. 2011; Aharonian et al. 2016). Thus, CX measurements offer an entirely new window toward understanding soft X-ray background emission, where the interaction between cold and hot gas flow is inaccessible to other observations (Dennerl 2010).

From a fundamental point of view, these CX collisions of astrophysical interest often occur at low collision energies with large total cross sections ($\sim 10^{-15}$ cm$^2$). The following X-ray emission has a high radiative decay rate and rich discrete spectral features etc., these unique features enable CX to be an important diagnostic entity for the astrophysical soft X-ray background in a hot astrophysical plasma ($\sim 10^6$–$10^8$ K) environment (McCammon et al. 2002; Kahn et al. 2005; Shull et al. 2012; Nicastro et al. 2018). Essentially, the CX-induced soft X-ray emission is determined by nl-resolved state-selective capture cross sections and favored by selection rules $\Delta l = \pm 1$ and cascade decay of the excited electrons ($n$ and $l$ are the principal quantum number and angular momentum quantum number, respectively). Experimentally, energy-gain spectroscopy (Okuno et al. 1983; Huber & Kahlert 1983; Schmeissner et al. 1984; Seredyuk et al. 2005), photon-emission spectroscopy (Hayakawa 1960; Bliman et al. 2002; Bodewits & Hoekstra 2007), cold target recoil-ion momentum spectroscopy (COLTRIMS; Abdallah et al. 1998; Kamber et al. 1999; Dörner et al. 2000; Fléchard et al. 2001; Fischer et al. 2002; Ullrich et al. 2003; Zhang et al. 2017), and COLTRIMS plus soft X-ray measurement (Ali et al. 2016; Hasan et al. 2001) have been used for measuring nl-resolved state-selective capture cross sections. However, the existing measurements are still sparse. Theoretically, the classical over-the-barrier model (COB) (Ryufuku et al. 1980; Burgdorfer et al. 1986), molecular coulombic barrier model (MCBM; Niehaus 1986), classical trajectory Monte Carlo (Olson & Salop 1976, 1977), multichannel Landau–Zener method (MCLZ; Cumbee et al. 2016), atomic orbital close coupling (Fritsch & Lin 1984), and molecular orbital close-coupling methods (Kimura & Lane 1989) along with quantum molecular orbital close coupling (Liu et al. 2016) etc. have been used for obtaining low-energy CX data. Another approach to obtain nl-resolved state-selective capture cross sections is the two-center basis generator method (TC-BGM; Leung & Kirchner 2016), which is a semiclassical, nonperturbative coupled-channel approach performed within the independent electron framework. However, such a calculation is challenging in terms of technique and computational resources.

Nevertheless, for simplicity, the X-ray astrophysical community relies on scaling laws, MCBM, and MCLZ etc. or data collecting through previous publications. By implementing these CX data into astrophysical models like AtomDB-ACX (Foster et al. 2012; Smith et al. 2014), SPEX-CX (Gu et al. 2015), SASAL (Liang et al. 2014) etc., soft X-ray spectra from astrophysical observations can be analyzed and assigned. For instance, Gu et al. (2016) proposed a novel scenario that the reported 3.5 keV line originally observed in several galaxy clusters (Boyarsky et al. 2014) can be explained by a set of $S^{15+}$ transitions from n   9 to the ground state following CX between $S^{16+}$ ions and neutral particles. Konami et al. (2011) and Cumbee et al. (2016) show that $Ne^{9+}$ X-ray emission due to CX is in excess of electron-impact excitation in the starforming galaxy M82. Krasnopolsky et al. (2002) identify that $Ne^{7+}$ X-ray emission is due to CX in comet McNaught-Hartley (C/1999 T1). However, CX data from these classical and semiclassical CX theories may not be sufficiently accurate. Moradmand et al. (2018) measured absolute single- and double-CX cross sections between $Si^{(7-10)+}$ and He and $H_2$ at solar wind energies and Hasan et al. (2001) measured relative SEC cross sections of $N^{7+}$ and $O^{7+}$ with CO, $CO_2$, and $H_2O$. Both measurements show that only limited agreement is found with the results calculated using the COB model. Betancourt- Martinez et al. (2018) present a high-resolution measurement of the L-shell CX of Ni ions and neutral $H_2$ and He, which shows the significant differences between experimental results and commonly used CX models. Thus, laboratory measurements are important in providing "reality" in extensive test cases that the level of confidence in the theoretical data has been regarded to be established.

More recently, significant progress in CX studies has been made with high-resolution X-ray microcalorimeters (Defay et al. 2013; Fogle et al. 2014; Seely et al. 2017; Betancourt- Martinez et al. 2018; Zhang et al. 2019) and crystal spectrometers (Beiersdorfer et al. 2005, 2018, 2019; Lepson et al. 2017). For instance, by measuring Lyβ/Lyα line ratios for X-ray emission in $C^{6+}$–He electron capture at solar wind velocities, Defay et al. (2013) clarified n  =  3 capture being dominant, which later on was correctly reproduced by Cumbee et al. (2017), while in a previous solar wind model Koutroumpa (2007) only assumed the dominant capture. Hell et al. (2016) pointed out that high-resolution measurement of Kα X-ray emission line energies in $Si^{(4-12)+}$ and $S^{(6-14)+}$ can be applied to redetermine the bulk motion of material in astrophysical sources. However, high-resolution X-ray measurements are inefficient in producing a CX database. On the other hand, it is really difficult to encounter the triple coincidence of high-resolution X-rays, scattered projectiles, and target recoil ions. The X-ray emission from singly excited ions following autoionizing double-electron capture cannot be distinguished from the SEC (Hasan et al. 2001; Seely et al. 2017), which results in a softening of the spectrum due to the contribution of  the autoionizing double-electron capture. Therefore, experimental and theoretical advances in obtaining nl-resolved state-selective cross sections is increasingly important for precise measurement and high-fidelity modeling.

In this paper, we report systematic measurements of  *n*-resolved state-selective capture cross sections for $Ne^{(8,9)+}$ CX with He and $H_2$ by using a COLTRIMS apparatus, since  $Ne^{(8,9)+}$ ions are minor highly charged ions in solar wind ions,  and He and $H_2$ are important components of the solar system  (Moradmand et al. 2018). The present pilot COLTRIMS  measurements aim to provide basic data on state-selective  capture cross sections and test the validity of CX theory.  Importantly, based on the state-selective capture cross section data measured by the state-of-the-art COLTRIMS, soft X-rays  following CX could be calculated by using l distributions and  the well-established radiative cascade theory; this may largely  minimize the data uncertainty for modeling CX X-ray emission  in astrophysics.

## 2.  EXPERIMENTAL SETUP

A new compact beam line has been built for highly charged ion–atom collisions at the Institute of Modern Physics, Chinese Academy of Sciences (IMP-CAS) in Lanzhou (Zhu et al. 2019). As shown in Figure 1, it consists of a Dresden electronbeam ion source (EBIS) and COLTRIMS detecting system, which largely extends the collision-energy regime of the existing 320 kV platform for multidisciplinary research with highly charged ions (Ma et al. 2009) to a few hundred eV/u. This facilitates the laboratory simulation of astrophysically relevant CX state selectivity in slow highly charged ion collisions with He and $H_2$. The present CX measurements were performed with the COLTRIMS at EBIS platform, except for these measurements involving 15.75 and 24.75 keV/u $Ne^{9+}$, for which we use the COLTRIMS at 320 kV platform. The following introduction about this facility was given by taking $Ne^{8+}$–He SEC as an example, which is described as

$$Ne^{8+} + He \rightarrow Ne^{7+}(1s^2 nl) + He^+(1s) \qquad (1)$$

In brief, this EBIS consists of a highly emissive metal-alloy cathode, an ion-trap assembly with a quadruple lens and magnetic field, electron collector, repeller, and ion optics. Its working principle is based on electron-impact ionization of neutral atoms at the ion-trap section and details can be found elsewhere (Schmidt et al. 2009, 2014). Ne gas was injected into the EBIS ion-trap section through a capillary. An electron beam emitted from the cathode was first compressed into an electron current density of more than 230 A cm$^{-2}$ by an axial magnetic field and then impact ionization of Ne gas to produce $Ne^{(1-10)+}$ ion series. The $Ne^{8+}$ ions were selected by a Wien filter according to the $V = \frac{E}{B}$ velocity selection relationship ($\vec{V}$ is the velocity of the projectile ion. $\vec{E}$ and $\vec{B}$ are electric field and magnetic field of the Wien filter, respectively), and then accelerated or decelerated to the desired energy. An einzel lens, two deflectors, and a slit are installed on the beam line for optimizing ion beams. Before entering into an ultrahigh vacuum (UHV) chamber, the $Ne^{8+}$ ion beam current is adjusted to be few picoamps and the beam size is less than 1 mm in diameter.

The working principles of the COLTRIMS detecting system were reported elsewhere (Dörner et al. 2000; Ullrich et al. 2003; Ma et al. 2011; Zhang et al. 2016, 2017; Guo et al. 2017). It mainly consists of a supersonic gas-jet target, TOF spectrometer, and projectile ion analyzer. He gas was cooled through adiabatic expansion in a two-stage differential pumping system under the driven pressure of 4 bar, and finally formed a locally high-density supersonic gas target inside the UHV chamber. The estimated target density is around $10^{10}$ atoms cm$^{-3}$. The formed He target collides with the $Ne^{8+}$ ion beam at a right angle in the center of the TOF spectrometer. The TOF spectrometer's electric field of 3 V cm$^{-1}$ was used to guide the recoil ions from the small overlap volume of the gas jet with the projectile beam onto a position-sensitive detector. The lengths of the acceleration section and the drift section were 107.5 and 215 mm, respectively, which meets the Wiley– McLaren time-focusing condition (Wiley & McLaren 1955). Charge-changed projectile ions were first analyzed by an electric deflector immediately downstream of the TOF spectrometer and then detected by another position-sensitive detector. The primary ion beam was collected by a Faraday cup.

During the measurement, the pressure of the UHV chamber was kept at about $1\times10^{-8}$ mbar. Coincidence measurement techniques were used through standard electronics. From the two-dimensional coincidence spectrum of the positions of the projectile ions on the position-sensitive detector and flight times of recoil ions, double-electron capture, SEC, and autoionization double-electron capture reaction channels can be identified. With the registered position and TOF information, the three-dimensional momentum of the recoil ions can be constructed. Here, a Cartesian coordinate was used, the measured momentum component along the beam direction was defined as the longitudinal momentum of recoil ions $P_{z\pi}$, and the square root of the sum of the other two momentum components was defined as the transverse momentum of recoil ions (Zhang et al. 2017). According to the principle of energy and momentum conservation, the state selectivity can be reflected by the measured longitudinal recoil-ion momentum. The dynamical relationship between the longitudinal momentum of recoil ions, the projectile velocity, and the binding energy of the active electron is shown in the following formula (Cassimi et al. 1996; Dörner et al. 2000; Ullrich et al. 2003):

$$P_{zr} = -\frac{Q}{V_p} - \frac{1}{2}V_p \tag{2}$$

where $v_p$ is the projectile velocity and Q is the binding energy difference of the active electron before and after collision and is defined as Q= $\varepsilon_i$ - $\varepsilon_f$. $\varepsilon_i$ and $\varepsilon_f$ are binding energies of the active electron in the initially ground-state target and finally excited ion, respectively. From Equation (2) it is straightforward to derive the following result:

$$Q = -\frac{1}{2}V_p^2 - V_p \cdot P_{zr} \tag{3}$$

For the initially ground-state He and $H_2$, values of the initial binding energy $\varepsilon_i$ are 24.6 eV and 15.4 eV, respectively. The binding energies of finally excited $Ne^{7+}$ and $Ne^{8+}$ via CX are available from the NIST Atomic Spectra Database (Kramida et al. 2019). Therefore, state selectivity in CX collisions can be easily identified by measuring the longitudinal momentum of recoil ions.

## 3. RESULTS AND DISCUSSION

### 3.1. n-resolved state-selective capture cross sections in charge exchange

Figure 2 shows the measured Q-value in the SEC process of $Ne^{8+}$–He and $H_2$ collisions. It is shown that the electron is mainly captured into a state with the principal quantum number n = 4 for He and n = 5 for $H_2$. This is in good agreement with previous results (Bliman et al. 1997; Bonnet et al. 1985; Roncin et al. 1986; Folkmann et al. 1989; Langereis et al. 1997; Fischer et al. 2002; Zhang et al. 2019). In theory, a simple scaling law on the n-state preference was introduced by Otranto et al. (2006), Palchikov & Shevelko (1995), and Janev & Winter (1985); the most probable principal quantum number *n* for capture is

$$n = \left(\frac{I_H}{I_t}\right)^{1/2} q^{3/4} \tag{4}$$

where It is the target's first ionization energy, $I_H$ is the first ionization energy of the hydrogen atom, and q is the charge state of the projectile. This scaling law predicts that electrons are mainly captured into n = 3.5 and n = 4.5 for He and $H_2$, respectively. Due to the smaller binding energy of electrons in $H_2$ than He, they are preferentially captured into a higher n state for $H_2$ than He. For the present energy range, with the increase of collision energy n = 3, 5 capture for He and n = 4, 6 capture for $H_2$ coming into play, this could be understood as more channels being open when the collision energies are increasing (Janev & Winter 1985).

Figure 3 shows the measured state selectivity of $Ne^{9+}$-He and $H_2$ electron capture. We found that electron is mainly captured into states of n = 4 for He below 6.75 keV/u, and contribution of n = 5 capture competitively increases above 6.75 keV/u. For $H_2$ target, electron is mainly captured into states of n = 5 and n = 6, these two captures have comparable intensities. Our measurements agree with previous results (Vernhet et al. 1985; Kimura et al. 1987; Folkmann et al. 1989; Lyons et al. 2017) for main electron capture channel. It is clearly shown the strong dependence of relative cross section on target ionization energy, i.e., SEC with low target ionization energy produces high capture state n, this can also be qualitatively predicted by the scaling law.

With the decrease of collision energy, it is anticipated that the electron capture becomes very selective for $Ne^{8+}$ and tends to converge to the only dominate channel, as suggested by Hasan et al. (2001), but this is not the case for $Ne^{9+}$ collision. Taking the He target as an example, n = 4 is predominant for the case of $Ne^{8+}$ collision, while additional n = 5 capture occurs at a considerable intensity compared with n = 4 capture for the case of $Ne^{9+}$ collision. A similar trend can also be found for the $H_2$ target. This indicates that the projectile charge state significantly affects state selectivity in the SEC collision process.

It is known that the longitudinal recoil-ion momentum resolution is mainly limited by the width of the supersonic target gas jet and that the gas-jet density profile is approximately described by a Gaussian profile. Therefore, the Gaussian curve fitting method was adopted to extract the relative state-selective capture cross sections, as shown in the right column of Figure 2, by taking 8 keV/u $Ne^{8+}$–$H_2$ capture as an example. Table 1 shows the measured relative state-selective capture cross sections of $Ne^{8+}$ with He and $H_2$, together with the MCLZ calculations obtained from the automated package Kronos (Mullen et al. 2016) and recent publications (Cumbee et al. 2016; Lyons et al. 2017). Stateselective capture cross sections of $Ne^{8+}$ with $H_2$ are calculated with the MCLZ theory by P. C. Stancil (2020, private communication). For dominant capture, we find that MCLZ agrees well with our measurements for $Ne^{8+}$ ion collisions with a He target for the present collision energies. There is a better than ~86% agreement between MCLZ and our measurements for the He target at present energies. For the $H_2$ target, the MCLZ calculations marginally agree with the measurements. For nondominant capture, the measured fraction of the sum of n = 3 and 5 capture is less than 10% for the He target, and for n = 4 and 6 capture it is less than 22% for the $H_2$ target.

Table 2 shows the measured relative electron capture cross sections of $Ne^{9+}$ with He and $H_2$. MCLZ calculations are in a better than ~55% agreement with the measurements for the dominant capture channel with the He target except for the collision energy of 24.75 keV/u. MCLZ does not adequately reproduce the relative state-selective capture cross sections of CX between $Ne^{9+}$ and $H_2$.

### 3.2. Soft X-ray emission following charge exchange

The soft X-ray emission following CX strongly depends on the distribution of angular momentum states populated by the CX process. For a given n capture, several analytical models for estimating the l population are approximated as a function of n, l, and q, which are widely used in the astrophysical community. They are the so-called statistical, separable, Landau–Zener I (LZ-I), Landau–Zener II (LZ-II), and even models (Smith et al. 2014; Gu et al. 2016). We have implemented these five l distributions for calculating soft X-ray spectra with our developed Photo Emission following Charge eXchange (PhECX) code in this work. Through n-resolved state-selectivity studies, we know that the n = 3, 4, 5 states are populated in CX between $Ne^{8+}$ and He for the present energy range. Soft X-ray emission following these captures only lies in the energy range from 100–220 eV. For the reason of convenience in our calculations, we use the statistical model as an example, showing the details of soft X-ray spectrum calculations. First, the relative cross sections of each n capture were obtained by normalizing to the sum of all measured n capture intensities, seen in Equation (5). The relative nl state capture cross sections were calculated by multiplying the statistical l distribution, shown in Equation (6). Second, the spectra of dipole allowed $3s \rightarrow 2p$, $3d \rightarrow 2p$, $3p \rightarrow 2s$, $4s$

$\rightarrow 2p$, $4d \rightarrow 2p$, $4p \rightarrow 2s$, $5s \rightarrow 2p$, $5d \rightarrow 2p$, and $5p \rightarrow 2s$ emissions to be calculated according to the branching ratios of the Ne$^{7+}$ cascade decay (Lindgard & Nielsen 1977); details are listed in Equations (7)–(15). Third, Equation (16) shows that the calculated intensity of each X-ray line is normalized to the sum of the soft X-ray spectrum. Here, the calculated soft X-ray spectrum was convoluted to the experimental resolution of 7.9 eV for comparison with the measurement by Zhang et al. (2019), which was corrected by the reported filter efficiency.

$$\sigma_n^{rel} = \frac{\sigma_n}{\sum_{n=3}^{5} \sigma_n} \tag{5}$$

$$\sigma_{nl}^{rel} = \left(\frac{2l+1}{n^2}\right) \cdot \sigma_n^{rel} \tag{6}$$

$$\sigma_{em}(3s \rightarrow 2p) = \sigma_{3s}^{rel} + 0.20 \cdot \left(\sigma_{4p}^{rel} + 0.22 \cdot \sigma_{5s}^{rel} + 0.09 \cdot \sigma_{5d}^{rel}\right) + 0.19 \cdot \sigma_{5p}^{rel} \tag{7}$$

$$\sigma_{em}(3p \rightarrow 2s) = \sigma_{3p}^{rel} + 0.41 \cdot \left(\sigma_{4s}^{rel} + 0.07 \cdot \sigma_{5p}^{rel}\right) + 0.24 \cdot \left(\sigma_{4d}^{rel} + 0.03 \cdot \sigma_{5p}^{rel} + 0.36 \cdot \sigma_{5f}^{rel}\right) + 0.23 \cdot \sigma_{5d}^{rel} + 0.31 \cdot \sigma_{5s}^{rel} \tag{8}$$

$$\sigma_{em}(3d \rightarrow 2p) = \sigma_{3d}^{rel} + 0.04 \cdot \left(\sigma_{4p}^{rel} + 0.22 \cdot \sigma_{5s}^{rel} + 0.09 \cdot \sigma_{5d}^{rel}\right) + \sigma_{4f}^{rel} + \sigma_{5g}^{rel} + 0.02 \cdot \sigma_{5p}^{rel} + 0.64 \cdot \sigma_{5f}^{rel} \tag{9}$$

$$\sigma_{em}(4s \rightarrow 2p) = 0.59 \cdot \left(\sigma_{4s}^{rel} + 0.07 \cdot \sigma_{5p}^{rel}\right) \tag{10}$$

$$\sigma_{em}(4p \rightarrow 2s) = 0.76 \cdot \left(\sigma_{4p}^{rel} + 0.22 \cdot \sigma_{5s}^{rel} + 0.09 \cdot \sigma_{5d}^{rel}\right) \tag{11}$$

$$\sigma_{em}(4d \rightarrow 2p) = 0.76 \cdot \left(\sigma_{4d}^{rel} + 0.03 \cdot \sigma_{5p}^{rel} + 0.36 \cdot \sigma_{5f}^{rel}\right) \tag{12}$$

$$\sigma_{em}(5s \rightarrow 2p) = 0.47 \cdot \sigma_{5s}^{rel} \tag{13}$$

$$\sigma_{em}(5p \rightarrow 2s) = 0.68 \cdot \sigma_{5p}^{rel} \tag{14}$$

$$\sigma_{em}(5d \rightarrow 2s) = 0.68 \cdot \sigma_{5d}^{rel} \tag{15}$$

$$\sigma_{em}^{rel}(nl \rightarrow 2l') = \frac{\sigma_{em}(nl \rightarrow 2l')}{\sum_{n=3}^{5} \sigma_{em}(nl \rightarrow 2l')} \tag{16}$$

Here $\sigma_n$ is the $n$-resolved state-selective cross section measured in our experiment, $\sigma_n^{rel}$ is the relative cross section of n capture, $\sigma_{em}(nl \rightarrow 2l')$ represents the emission cross section where the coefficients on the right side of Equations (7)–(15) are branching ratios of cascade decay (Lindgard & Nielsen 1977; Politis et al. 1987), and $\sigma_n^{rel}$ is the normalized emission cross section. Similarly, the normalized soft X-ray spectrum can also be calculated for the separable, LZ-I, LZ-II, and even models. Since the fraction of n = 5 capture is only 1.5%, its contribution to 4d, 4p population by cascade can be neglected as well as the emission decaying to n = 2.

We calculated the spectra and line intensities of soft X-ray cascade emissions following 4 keV/u Ne$^{8+}$ CX with He, which are correspondingly in the range of the fast solar wind velocities. It should be pointed out that the MCLZ method explicitly predicts $nl$-resolved cross sections without $l$ distribution (Lyons et al. 2017). In SPEX-CX, the Ne$^{7+}$ spectrum is calculated in the most empirical way, i.e., a total cross section and the $n$ and $l$ populations were calculated with Equation (2), Equation (3), and the LZ-I scaling equation from Gu et al. (2016), respectively. Figure 4 shows the comparison between the soft X-ray spectrum from various calculations and the measured spectrum, which are split into two graphs for a better visualization.

Quantitatively, the fractional emission cross sections are given in Table 3. Compared with the measurement, it is clear that the SPEX-CX model overestimates the contribution from the $5p \rightarrow 2s$ transition and underestimates the contribution of $4d \rightarrow 2p$ and $4s \rightarrow 2p$. The agreement between the even model calculations and measurements is better than 90% for $4d \rightarrow 2p$ and $4p \rightarrow 2s$ and around 78% for $4s \rightarrow 2p$ transitions, while it is 66% for $3d \rightarrow 2p$ and 62% for $3p \rightarrow 2s$ transitions. For separable model calculations, the agreement with measurements is around 90% for $3d \rightarrow 2p$, 82% for $4s \rightarrow 2p$, and 81% for $4d \rightarrow 2p$ transitions, while it is only 69% for $4p \rightarrow 2s$ and 55% for $3p \rightarrow 2s$ transitions, respectively. For the calculations with statistical and LZ-I distributions, they significantly deviate from the measurement for $3d \rightarrow 2p$ and $4p \rightarrow 2s$, respectively. A possible reason is that the collision energy is too high for LZ-I and too low for the statistical model. Since their energy range of validity is not robustly known for these l distributions, the nature of CX velocity dependence is not taken into consideration. Also, two semiclassical approaches are available for comparison with the measurement. One is the TC-BGM, in which the time dependent screening potential was used to model the target response during the collision. The calculation of the TC-BGM significantly deviates from the measurement except $4d \rightarrow 2p$ and $4s \rightarrow 2p$ transitions. The other one is MCLZ, which is in fact a dynamical calculation including rotational couplings. For MCLZ calculations, the agreement with the measurement is better than 90% for $3d \rightarrow 2p$ and $4d \rightarrow 2p$, about 88% for $4s \rightarrow 2p$, 70% for $4p \rightarrow 2s$, and 63% for $3p \rightarrow 2s$ transitions. Overall, the MCLZ calculation is in a better agreement with the measurement. This makes sense in that the average orbital velocity of the active target electron is mostly larger than the collision velocity for the present collision system, and electrons adjust to the motion of nuclei from both collision partners through dynamical couplings. Therefore, more comparisons between measurements and MCLZ calculations are necessary for providing accurate l-resolved state-selective capture between slow highly charged ions and neutrals in astrophysical plasma.

## 4. SUMMARY

With the well-developed COLTRIMS techniques, a series of measurements of n-resolved electron capture between $Ne^{(8,9)+}$ and He and $H_2$ have been carried out. The relative contribution of each n-resolved state-selective capture has been obtained. By comparison with the available MCLZ calculations, which are often used for modeling diffusive X-ray emission in astrophysical contexts, we found that the results from MCLZ are in a good agreement with our measurements for $Ne^{(8,9)+}$ CX with He, while they only generally agree with the measurements for $Ne^{(8,9)+}$ CX with $H_2$. Additionally, with our n-resolved measurements, we calculate the soft X-ray emissions following CX between 4 keV/u $Ne^{8+}$ and He for testing the reliability of these often-used l distributions and MCLZ calculations. It is found that soft X-ray spectrum features can be reproduced by separable and even model calculations, while there are some significant deviations from the measurement for other l distributions. For the soft X-ray spectrum calculated with MCLZ calculations, the results are in a better agreement with the measurement.

Our apparatus is being modified for the high-resolution measurement of $nl$ state-selective electron capture cross section measurement. These modifications are not only necessary for testing CX theory, but also can provide accurate data for modeling the soft X-ray emission background in astrophysical plasma environment.

## 5. Acknowledgments

This work was supported by the National Key Research and Development Program of China (grant Nos. 2017YFA0402400 and 2017YFA0402300), Strategic Key Research Program of the Chinese Academy of Sciences (XDB34020000), and the National Natural Science Foundation of China (grant No. U1832201). The authors are grateful to Professor P. C. Stancil for providing us with the MCLZ calculations and Dr. Anthony C. K. Leung for providing us with the TC-BGM calculations, which made this work possible. We also thank the engineers of the EBIS platform for providing us with the high-quality ion beam and for their assistance during the experiment.

**Figure 1**. Top view of the CX experiment with the EBIS beam line and COLTRIMS setup at IMP-CAS; the supersonic gas-jet flow direction is from bottom to top. $E_{TOF}$ represents the electric field of the time-of-flight (TOF) spectrometer.

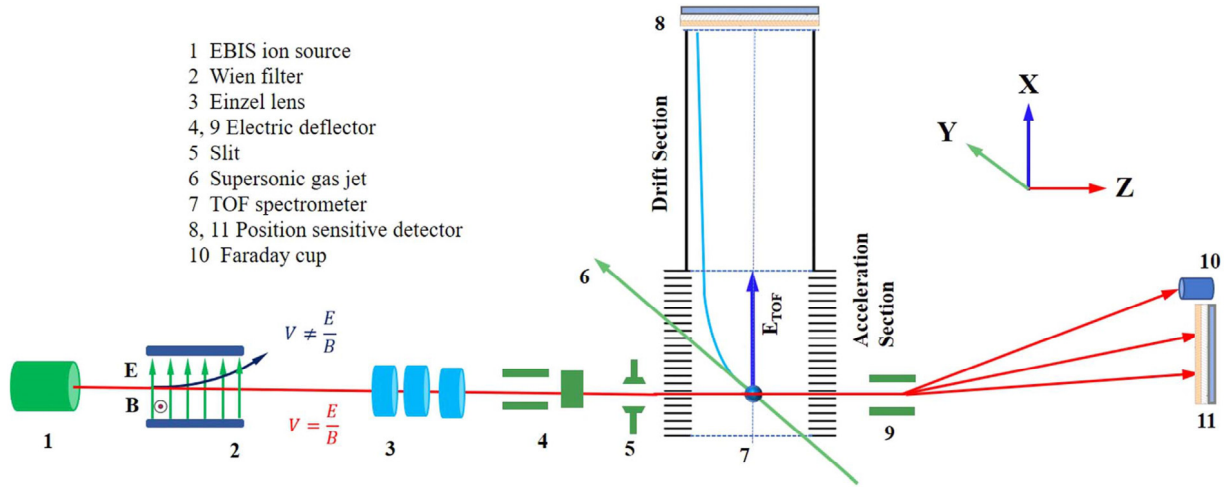

1 EBIS ion source
2 Wien filter
3 Einzel lens
4, 9 Electric deflector
5 Slit
6 Supersonic gas jet
7 TOF spectrometer
8, 11 Position sensitive detector
10 Faraday cup

**Figure 2.** The measured Q spectra of SEC in Ne$^{8+}$ collisions with He and H$_2$. The left and right panels represent state-selective capture cross sections for He, H$_2$ target, respectively. The incident projectile energies of 1, 2, 4, 6, and 8 keV/u from up to down, respectively. Hollow black squares are experimental measurements and the red lines are used to guide for the eyes. The blue lines in the bottom-right panel represent Gaussian curve fitting (see the text).

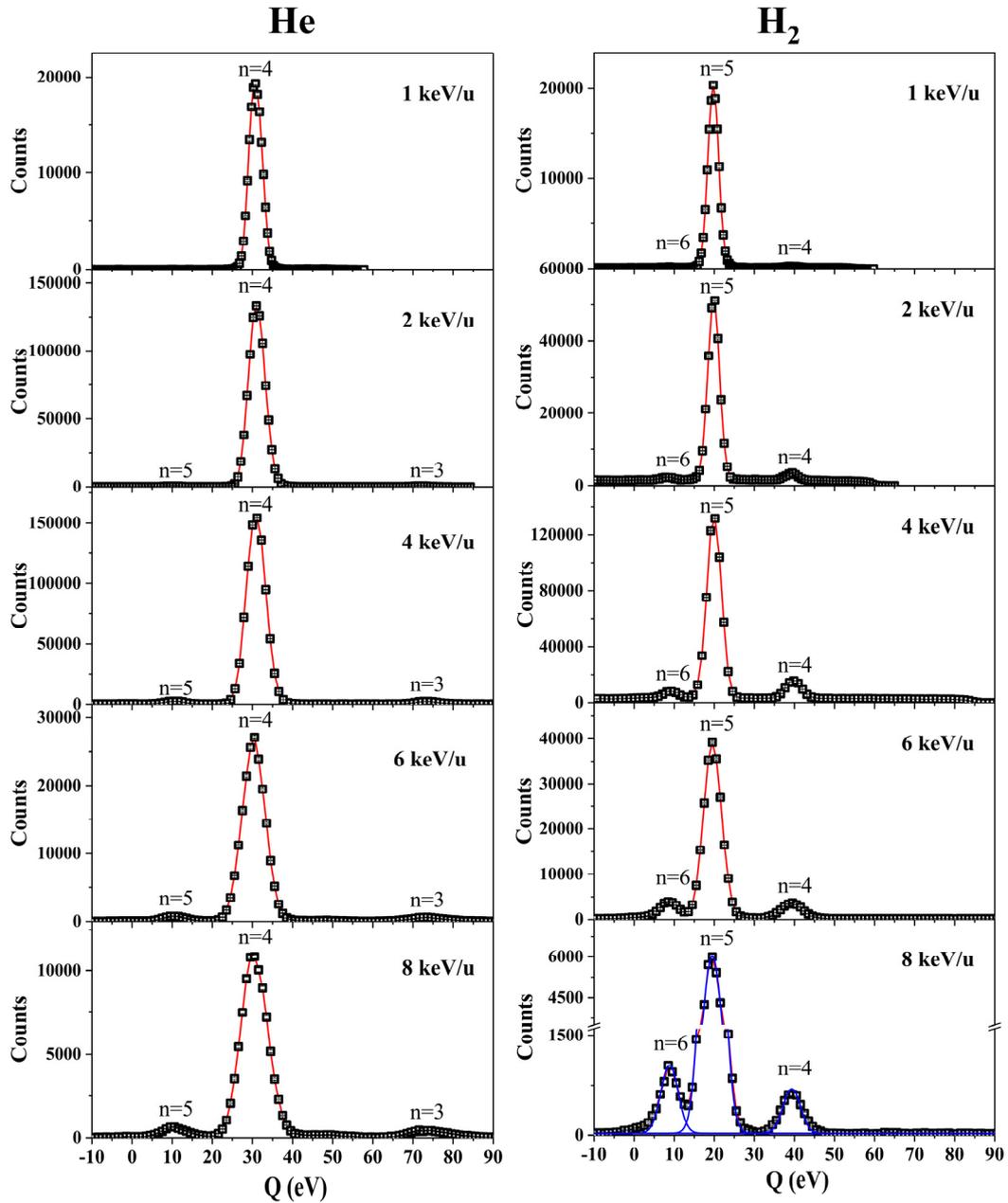

**Figure 3.** The measured Q spectra of single electron capture between Ne$^{9+}$ and He and H$_2$. The left and right columns represent He and H$_2$ targets, respectively. Panels from top to bottom show the incident projectile energies of 2.25, 4.5, 6.75, 15.75, and 24.75 keV/u, respectively. Hollow black squares are experimental measurements and the red lines are to guide the eyes.

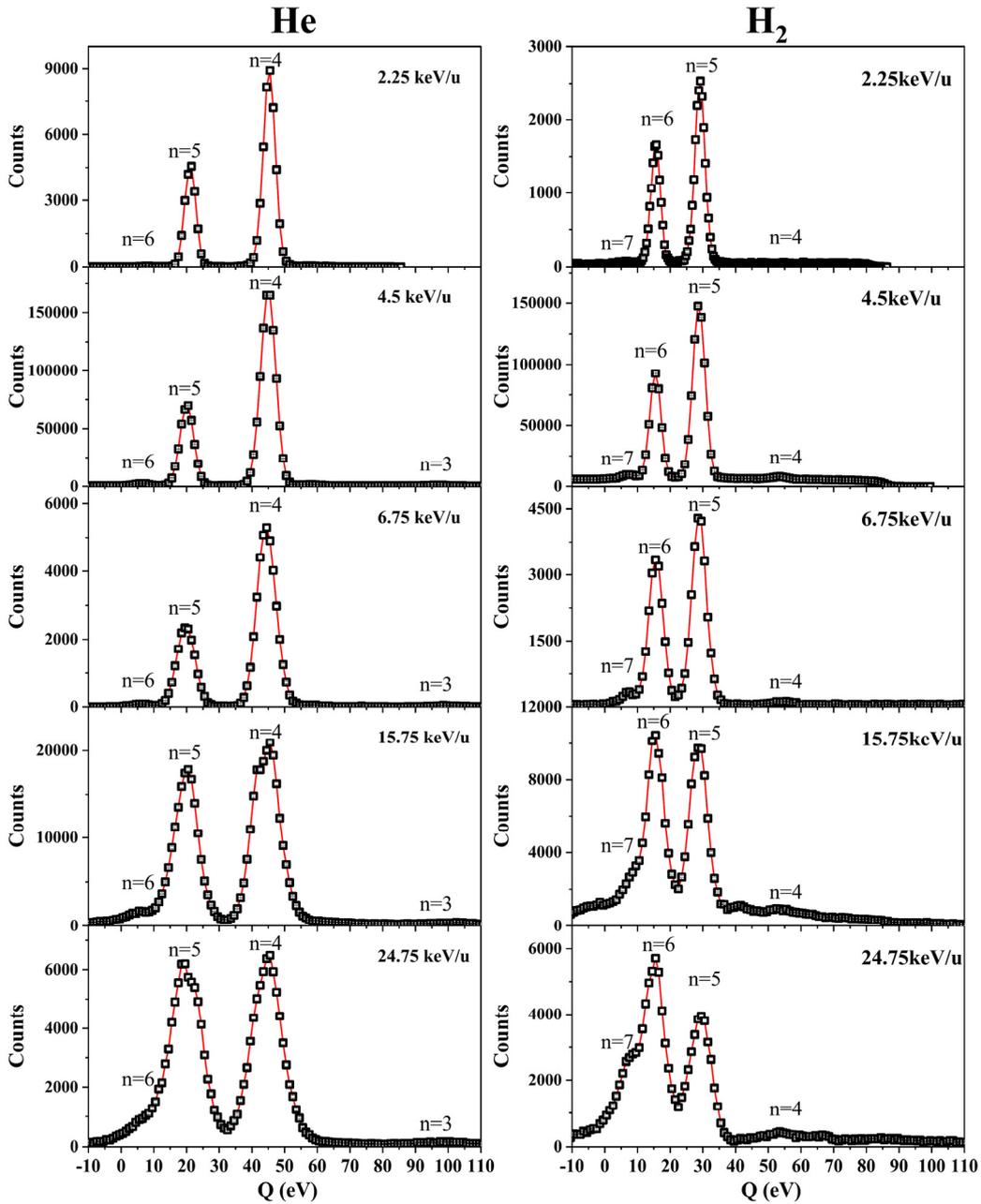

**Figure 4.** The normalized emission spectra of Ne$^{7+}$ following CX between 4 keV/u Ne$^{8+}$ and He. The black, red, green, blue and magenta solid lines represent statistical, separable, Landau–Zener I (LZ-I), Landau–Zener II (LZ-II), and even models, respectively. The purple and blue dashed lines are the results calculated by MCLZ and TC-BGM, respectively, the orange line is the result calculated by SPEX-CX, and the black filled squares represent recent measurements by Zhang et al. (2019).

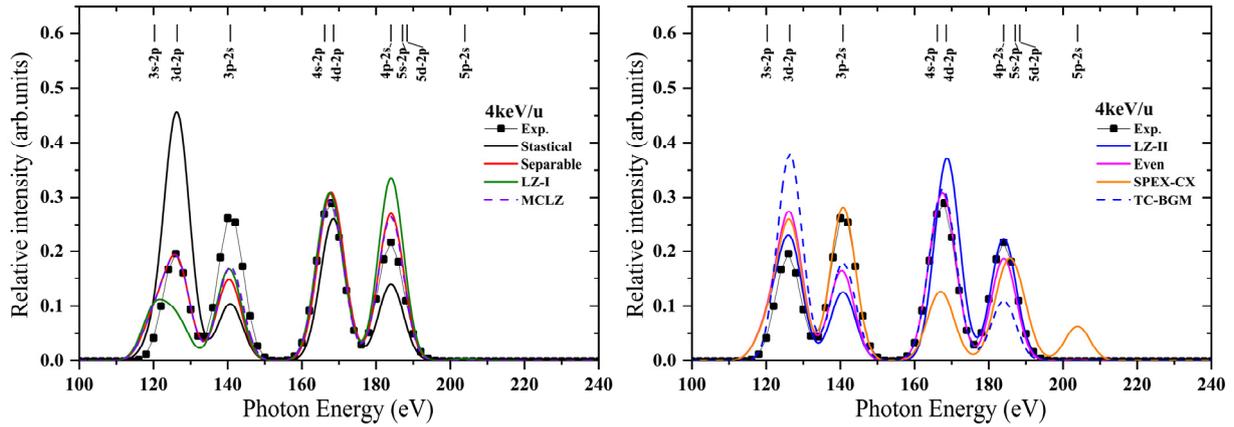

**Table 1**
Comparison of Measured and Calculated Relative Cross Sections of the *n*-Resolved State-selective SEC Process in Ne$^{8+}$ Collisions with He and H$_2$

| Energy (keV/u) | He | | | H$_2$ | | |
| --- | --- | --- | --- | --- | --- | --- |
| | n | $\sigma_{Exp}^{rel}$ (%) | $\sigma_{MCLZ}^{rel}$ (%) | n | $\sigma_{Exp}^{rel}$ (%) | $\sigma_{MCLZ}^{rel}$ (%) |
| 1 | 3 | | 5.60 | 3 | | 0.13 |
| | 4 | 100 | 94.39 | 4 | 0.42(1.91) | 35.23 |
| | 5 | | 0.01 | 5 | 98.55(0.69) | 64.61 |
| | 6 | | | 6 | 0.39(2.07) | 0.03 |
| 2 | 3 | 0.42(0.19) | 9.08 | 3 | | 0.65 |
| | 4 | 99.25(1.27) | 90.91 | 4 | 4.86(0.72) | 43.39 |
| | 5 | 0.33(0.21) | 0.01 | 5 | 92.89(0.80) | 55.93 |
| | 6 | | | 6 | 2.25(0.52) | 0.03 |
| 4 | 3 | 1.56(0.17) | 13.57 | 3 | | 1.91 |
| | 4 | 96.98(0.24) | 86.42 | 4 | 10.79(0.62) | 50.79 |
| | 5 | 1.46(0.16) | 0.008 | 5 | 84.88(0.63) | 47.28 |
| | 6 | | | 6 | 4.33(0.44) | 0.02 |
| 6 | 3 | 3.06(0.21) | 16.52 | 3 | | 3.17 |
| | 4 | 94.27(0.36) | 83.47 | 4 | 8.73(0.35) | 54.84 |
| | 5 | 2.67(0.18) | 0.007 | 5 | 82.52(0.30) | 41.97 |
| | 6 | | | 6 | 8.75(0.18) | 0.02 |
| 8 | 3 | 5.02(0.35) | 18.91 | 3 | | 4.37 |
| | 4 | 90.21(0.52) | 81.08 | 4 | 8.76(0.35) | 57.46 |
| | 5 | 5.02(0.28) | 0.006 | 5 | 78.23(0.62) | 38.15 |
| | 6 | | | 6 | 13.01(0.38) | 0.02 |

Note. Errors given in the brackets are statistical errors which are determined from the Gaussian fitting of the peak. $\sigma_{Exp}^{rel}$ and $\sigma_{MCLZ}^{rel}$ represent the measured relative cross section and the relative cross section calculated with MCLZ, respectively.

**Table 2**
Comparison of Measured and Calculated Relative Cross Sections of the *n*-Resolved State-selective SEC Process in Ne$^{9+}$ Collisions with He and H$_2$

| Energy (keV/u) | He | | | H$_2$ | | |
| --- | --- | --- | --- | --- | --- | --- |
| | n | $\sigma_{Exp}^{rel}$ (%) | $\sigma_{MCLZ}^{rel}$ (%) | n | $\sigma_{Exp}^{rel}$ (%) | $\sigma_{MCLZ}^{rel}$ (%) |
| | 3 | | 1.09 | 3 | | 3.72 |
| 2.25 | 4 | 67.70(0.47) | 80.26 | 4 | 61.29(0.90) | 79.34 |
| | 5 | 32.11(0.36) | 19.57 | 5 | 37.78(0.75) | 16.93 |
| | 6 | 0.19(0.14) | 2.70$^{-7}$ | 6 | 0.94(0.29) | 5.86$^{-4}$ |
| | 3 | 0.30(0.10) | 2.67 | 4 | 1.19(0.84) | 7.41 |
| 4.5 | 4 | 71.63(0.17) | 84.79 | 5 | 62.36(0.79) | 80.04 |
| | 5 | 27.21(0.13) | 13.81 | 6 | 34.25(0.83) | 12.52 |
| | 6 | 0.87(0.10) | 1.94$^{-7}$ | 7 | 2.21(0.95) | 4.30$^{-4}$ |
| | 3 | 0.54(0.20) | 4.01 | 4 | 1.72(0.31) | 10.15 |
| 6.75 | 4 | 68.83(0.50) | 85.57 | 5 | 53.63(0.53) | 79.28 |
| | 5 | 29.63(0.37) | 11.29 | 6 | 40.97(0.51) | 10.48 |
| | 6 | 1.00(0.19) | 1.58$^{-7}$ | 7 | 3.68(0.31) | 3.58$^{-4}$ |
| | 3 | 0.85(0.42) | 7.75 | 4 | 1.90(0.81) | 16.60 |
| 15.75 | 4 | 54.65(0.43) | 85.57 | 5 | 44.56(1.10) | 75.42 |
| | 5 | 40.89(0.41) | 7.54 | 6 | 44.73(1.13) | 7.41 |
| | 6 | 3.61(0.26) | 1.02$^{-7}$ | 7 | 8.81(0.87) | 2.50$^{-4}$ |
| | 3 | 1.34(0.31) | 10.21 | 4 | 2.47(0.72) | 20.09 |
| 24.75 | 4 | 47.71(0.45) | 84.79 | 5 | 34.70(0.96) | 72.51 |
| | 5 | 43.82(0.38) | 6.17 | 6 | 43.84(1.02) | 6.25 |
| | 6 | 7.13(0.28) | 7.82$^{-8}$ | 7 | 18.99(0.81) | 2.10$^{-4}$ |

**Note.** Errors given in the brackets are statistical errors which are determined from the Gaussian fitting of the peak. $\sigma_{Exp}^{rel}$ and $\sigma_{MCLZ}^{rel}$ represent the measured relative cross section and the relative cross section calculated with MCLZ, respectively.

**Table 3**

Fraction of Soft X-Ray Emission Following CX between 4 keV/u Ne$^{8+}$ and He

| Line | Photon Energy (eV) | Statistical | Separable | LZ-I | LZ-II | Even | SPEC-CX | TC-BGM | MCLZ | Experiment |
|------|-----|-----|-----|-----|-----|-----|-----|-----|-----|-----|
| $3s \rightarrow 2p$ | 120.28 | 3.84 | 7.54 | 9.35 | 5.89 | 5.44 | 5.30 | | 7.53 | |
| $3d \rightarrow 2p$ | 126.36 | 44.79 | 17.69 | 6.97 | 21.69 | 26.22 | 24.90 | 37.80 | 17.66 | 19.57 |
| $3p \rightarrow 2s$ | 140.73 | 10.38 | 14.79 | 16.72 | 12.56 | 16.47 | 28.10 | 17.83 | 17.12 | 26.97 |
| $4s \rightarrow 2p$ | 166.12 | 3.59 | 9.66 | 14.35 | 0.01 | 14.31 | 8.28 | 13.64 | 13.16 | 11.72 |
| $4d \rightarrow 2p$ | 168.54 | 23.19 | 22.91 | 18.51 | 37.04 | 18.51 | 5.22 | 19.81 | 18.13 | 19.27 |
| $4p \rightarrow 2s$ | 184.00 | 13.87 | 26.76 | 33.30 | 22.19 | 18.50 | 11.36 | 10.92 | 26.41 | 20.37 |
| $5s \rightarrow 2p$ | 187.13 | 0.03 | 0.11 | 0.11 | | 0.13 | 5.68 | | | 1.58 |
| $5d \rightarrow 2p$ | 188.35 | 0.19 | 0.29 | 0.29 | 0.42 | 0.20 | 4.87 | | | 0.52 |
| $5p \rightarrow 2s$ | 203.92 | 0.12 | 0.34 | 0.34 | 0.19 | 0.30 | 6.28 | | | |

**Note.** These five *l* distributions are listed in Gu et al. (2016). TC-BGM results are from Leung & Kirchner (2019), MCLZ results are from Lyons et al. (2017), and the experimental measurements are from Zhang et al. (2019). The energy of the emission line is obtained from the NIST database (Kramida et al. 2019).